\documentclass[aps,prd,preprint,groupedaddress]{revtex4}
\usepackage{epsfig,dcolumn}
\usepackage{graphicx}
\usepackage{bm}
\newcommand{\nc}{\newcommand}
\nc{\mb}[1]{\makebox[#1]{}}
\nc{\CC}{{\scriptscriptstyle CC}}
\nc{\NC}{{\scriptscriptstyle NC}}
\nc{\V}{{\rm v}}
\nc{\SC}{{\rm s}}
\nc{\W}{{\scriptscriptstyle W}}
\nc{\X}{{\scriptscriptstyle X}}
\nc{\Z}{{\scriptscriptstyle Z}}
\nc{\CS}{{\scriptscriptstyle CS}}
\nc{\SV}{{\scriptscriptstyle SV}}
\nc{\DY}{{\scriptscriptstyle DY}}
\nc{\PW}{{\scriptscriptstyle PW}}
\nc{\SB}{{\scriptscriptstyle SB}}
\nc{\CSV}{{\scriptscriptstyle CSV}}
\nc{\QCD}{{\scriptscriptstyle QCD}}
\nc{\GLS}{{\scriptscriptstyle GLS}}
\nc{\CIB}{{\scriptscriptstyle CIB}}
\nc{\PT}{{\scriptscriptstyle PT}}
\nc{\ASYM}{{\scriptscriptstyle asym}}
\nc{\IE}{{\it i.e.,\ }}
\nc{\EG}{{\it e.g.,\ }}
\nc{\EA}{{\it et al.}}
\nc{\AH}{{\it ad hoc\ }}
\nc{\CHPT}{{$\chi_{\PT}$\ }}

\nc{\NCA}{{\em Nuovo Cimento}}
\nc{\NIM}{{\em Nucl. Instrum. Methods}}
\nc{\NIMA}{{\em Nucl. Instrum. Methods} A}
\nc{\NPB}{{\em Nucl. Phys.} B}
\nc{\PLB}{{\em Phys. Lett.}  B}
\nc{\PRL}{{\em Phys. Rev. Lett.}}

\nc{\PRD}{{\em Phys. Rev.} D}
\nc{\PRC}{{\em Phys. Rev.} C}
\nc{\ZPC}{{\em Z. Phys.} C}

\nc{\st}{\scriptstyle}
\nc{\sst}{\scriptscriptstyle}
\nc{\mco}{\multicolumn}
\nc{\epp}{\epsilon^{\prime}}
\nc{\vep}{\varepsilon}
\nc{\ra}{\rightarrow}
\nc{\ppg}{\pi^+\pi^-\gamma}
\nc{\xpi}{x_{\pi}}
\nc{\pis}{\pi_{\SC}(\xpi)}
\nc{\piv}{\pi_{\V}(\xpi)}
\nc{\pist}{\widetilde{\pi}_{\SC}(\xpi)}
\nc{\nuN}{{\nu N_0}}
\nc{\nubN}{{\overline{\nu} N_0}}
\nc{\ovnu}{{\overline{\nu}}}
\nc{\dvx}{{d_{\V}(x)}}
\nc{\deldv}{{\delta \dvx}}
\nc{\uvx}{{u_{\V}(x)}}
\nc{\qvx}{{q_{\V}(x)}}
\nc{\deluv}{{\delta \uvx}}
\nc{\bux}{{\bar{u}(x)}}
\nc{\bdx}{{\bar{d}(x)}}
\nc{\dub}{{\delta \bux}}
\nc{\ddb}{{\delta \bdx}}
\nc{\snuNC}{{\langle \sigma^{\nuN}_{\NC}\rangle }}
\nc{\snubNC}{{\langle \sigma^{\nubN}_{\NC}\rangle }}
\nc{\snuCC}{{\langle \sigma^{\nuN}_{\CC}\rangle }}
\nc{\snubCC}{{\langle \sigma^{\nubN}_{\CC}\rangle }}
\nc{\Rnu}{{R^{\nu}}}
\nc{\Rnub}{{R^{\overline{\nu}}}}
\nc{\sintW}{{\sin^2 \theta_{\W} }}
\nc{\MS}{{\overline{MS}}}
\nc{\vp}{{\bf p}}
\nc{\rz}{{\rho_0^2}}
\nc{\ko}{K^0}
\nc{\kb}{\bar{K^0}}
\nc{\al}{\alpha}
\nc{\ab}{\bar{\alpha}}
\nc{\be}{\begin{equation}}
\nc{\ee}{\end{equation}}
\nc{\bea}{\begin{eqnarray}}
\nc{\eea}{\end{eqnarray}}
\nc{\beast}{\begin{eqnarray*}}
\nc{\eeast}{\end{eqnarray*}}

\begin{document}

\titlepage

\title{Experimental Tests of Charge Symmetry Violation in Parton 
Distributions} 

\author{J.T.Londergan}

\email{tlonderg@indiana.edu}
\affiliation{Department of Physics and Nuclear
            Theory Center,\\ Indiana University,\\ 
            Bloomington, IN 47405, USA}

\author{D.P. Murdock}
\email{murdock@tntech.edu}
\affiliation {Department of Physics,\\ Tennessee Technological University\\ 
                Cookeville, TN 38505, USA}

\author{A.W.Thomas}
\email{awthomas@jlab.org}
\affiliation {Jefferson Lab, 12000 Jefferson Ave.,\\ 
                Newport News, VA 23606, USA}
\date{\today}


\begin{abstract} 
{}Recently, a global phenomenological fit to high energy data has 
included charge symmetry breaking terms, leading to limits on the 
allowed magnitude of such effects. We discuss two 
possible experiments that could search for isospin violation in 
valence parton distributions. We show that, given the magnitude of 
charge symmetry violation consistent with existing global data, such 
experiments might expect to see effects at a level of several percent. 
Alternatively, such experiments could significantly decrease the 
upper limits on isospin violation in parton distributions. 
\end{abstract}



\pacs{11.30.Hv, 12.15.Mm, 12.38.Qk, 13.15.+g}

\maketitle


\section{Introduction\label{Sec:Intro}}

Charge symmetry is a particular form of isospin 
invariance that involves a rotation of $180^\circ$ about the ``2'' axis 
in isospin space.  For parton distributions, charge symmetry involves 
interchanging up and down quarks while simultaneously interchanging 
protons and neutrons.  In nuclear physics, charge symmetry is 
an extremely well respected symmetry, generally 
obeyed at the level of a fraction of a percent~\cite{Miller,Henley}.  
Since charge symmetry is so well satisfied at lower energies, it is 
natural to assume that it holds for parton distribution functions 
(PDFs).  Furthermore, there is no direct experimental evidence that 
charge symmetry is violated in PDFs. Recently, the experimental 
upper limits on parton charge symmetry violation (CSV) have been improved, 
by comparing the $F_2$ structure functions for deep inelastic scattering 
induced by muons and neutrinos  
\cite{Sel97,Ama91,Arn97,Lon04}. Until now, all phenomenological PDFs have 
assumed the validity of charge symmetry. But a recent global fit of PDFs 
by Martin \EA~\cite{MRST03} included for the first 
time the possibility of charge symmetry violating PDFs for both 
valence and sea quarks.  This provides us with parton 
distribution functions that agree with all of the experimental 
information used to obtain global fits to PDFs and which   
incorporate isospin violation. 

The global fit of Martin {\it et al.} assumed a particular form for the 
charge symmetry violating PDFs, for both valence and sea quark 
CSV.  In Sect.\ \ref{Sec:Csymmph}, we will review the form for 
the CSV terms used by MRST, and we will discuss some of the features 
of their amplitudes.  The allowed variation in the CSV terms is rather 
large.  Consequently, there may be possibilities either to measure 
the magnitude of parton CSV distributions, or to provide more strict 
experimental limits on these quantities, with dedicated experiments that 
focus on specific observables.  We will review two such possibilities 
in this paper, both of which are sensitive to valence quark 
CSV.       

One promising approach would be to measure Drell-Yan cross sections 
induced by charged pions on an isoscalar target, for which the simplest 
would be the deuteron.  In Sect.\ \ref{Sec:dyCSV}, we review the 
possibilities for such measurements.  We describe the observable 
which would be most sensitive to valence parton CSV distributions, and we 
use existing parton distributions, plus the CSV terms of MRST, to 
show potential variations in these observables which correspond to 
the current limits on CSV obtained in the MRST global fit.  

A second possibility is through semi-inclusive deep inelastic 
scattering for electrons on isoscalar targets.  In Sect.\ 
\ref{Sec:SIDIS}, we describe the relevant observable and show 
the magnitude of the effects which one must measure in order to 
establish charge symmetry 
violation in parton distribution functions, or to provide more 
strict upper limits on CSV terms.  

We then review the information that could be extracted from 
these experiments, and the prospects for such measurements.    

\section{Phenomenological Charge Symmetry 
Violating PDFs\label{Sec:Csymmph}}

Because CSV effects are extremely small at nuclear 
physics energy scales \cite{Miller,Henley}, it is natural to 
assume that parton distribution functions obey charge symmetry.  
Furthermore, there is no direct evidence for violation of parton 
charge symmetry \cite{Lon98}, even though the existing direct 
upper limits on parton charge symmetry violation are at roughly the 
5-10\% level \cite{Lon04}. Martin, Roberts, Stirling and Thorne (MRST) 
\cite{MRST03} have studied the uncertainties in parton distributions arising 
from a number of factors, including for the first time isospin violation.  

Charge symmetry violating PDFs involve the difference between, say, 
the down quark PDF in the proton and the up quark in the neutron; 
thus we define   
\bea
  \deldv &\equiv& d_{\V}^p(x) - u_{\V}^n(x) \nonumber \\ 
  \deluv &\equiv& u_{\V}^p(x) - d_{\V}^n(x) \ ,  
\label{eq:CSVdef}
\eea
with analogous relations for antiquarks. Now, from valence quark 
normalization, the first moment of the valence quark CSV PDFs must 
vanish, \IE 
\be 
  \int_0^1 \,dx\,\deldv  = \int_0^1 \,dx\, \left[ d_{\V}^p(x) - u_{\V}^n(x)
  \right] = 0 \ ,
\label{eq:quarknorm}
\ee
with an analogous relation for the first moment of $\deluv$.    

The MRST group chose a specific model for valence quark charge 
symmetry violating PDFs.  They constructed a function that automatically 
satisfied the quark normalization condition of Eq.~(\ref{eq:quarknorm}), 
namely:
\bea 
 \delta u_{\V}(x) &=& - \delta d_{\V}(x) = \kappa f(x) \nonumber \\ 
  f(x) &=& (1-x)^4 x^{-0.5}\, (x - .0909)  \, .
\label{eq:CSVmrst}
\eea
The function $f(x)$ in Eq.~(\ref{eq:CSVmrst}) was chosen so that at both 
small and large $x, f(x)$ 
has a form similar to the MRST2001 valence quark distributions 
\cite{MRST01}, and the first moment of $f(x)$ is zero. The functional form 
of the valence CSV distributions guaranteed that $\delta u_{\V}$ and 
$\delta d_{\V}$ would have opposite signs at large $x$, in agreement with 
theoretical models for parton CSV \cite{Sat92,Rod94}.  Fixing the form 
of the CSV parton distribution leaves undetermined only the overall 
coefficient $\kappa$, which was then varied in a global fit to a wide 
range of high energy data.   

\begin{figure}
\includegraphics[width=2.2in]{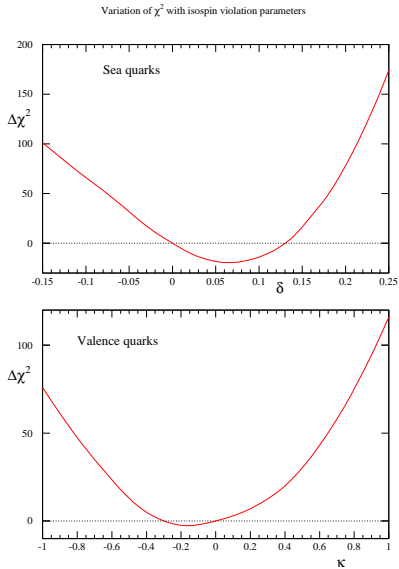}
\caption{$\chi^2$ profile for phenomenological isospin violating 
parton distributions, for sea quarks (top curve) and valence quarks 
(bottom curve), from the MRST 
group, Ref.\ \protect{\cite{MRST03}}. The quantity $\delta$ associated 
with sea quark isospin violation is defined  in Eq.\ 
(\protect{\ref{eq:seaCSV}}), while the coefficient $\kappa$ is defined in 
Eq.\ (\protect{\ref{eq:CSVmrst}}).
\label{Fig:chi2}}
\end{figure}

\begin{figure}[t]
\vspace{-1.0cm}
\hspace{-1.2cm} 
\includegraphics[width=2.2in]{MRSTcsv.eps}
\includegraphics[width=3.4in]{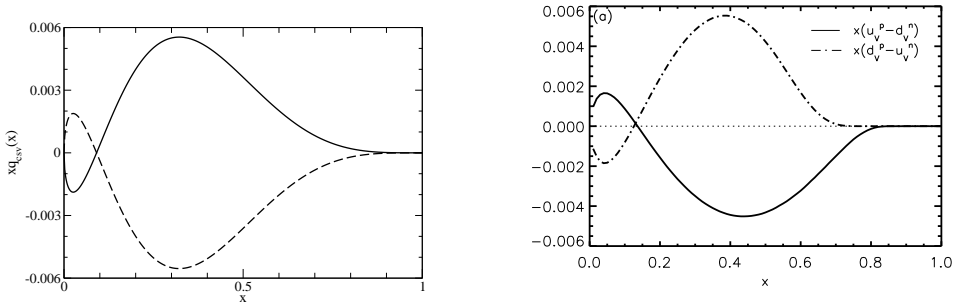}
\caption{Left: the phenomenological valence quark CSV function from Ref.\ 
\protect{\cite{MRST03}}, corresponding to best fit value $\kappa = -0.2$ 
defined in Eq.~(\protect\ref{eq:CSVmrst}). Solid curve: 
$x\deldv$; dashed curve: $x\deluv$. Right: theoretical 
CSV PDFs by Rodionov \EA, Ref. \protect\cite{Rod94}. Solid line: 
$x\deluv$; dash-dot line: $x\deldv$.  
\label{Fig:MRSTfx}}
\end{figure}

The value of $\kappa$ which minimised the $\chi^2$ in the MRST global 
fit was $\kappa = -0.2$.  The MRST $\chi^2$ vs.\ $\kappa$ is shown as 
the bottom curve in Fig.\ \ref{Fig:chi2}.  
Clearly $\chi^2$ has a shallow minimum with the 90\% confidence level 
obtained for the range $-0.8 \le \kappa \le +0.65$.  The top figure 
in Fig.~\ref{Fig:MRSTfx} plots the valence quark CSV PDFs corresponding 
to the MRST best fit value, $\kappa = -0.2$.  The best-fit phenomenological 
valence CSV PDFs look extremely similar to the CSV PDFs  
calculated by Rodionov {\it et al.} \cite{Rod94,Lon04}, who implemented 
charge symmetry violation in simple quark models; this is shown as 
the lower figure in Fig.~\ref{Fig:MRSTfx}.  Note, however, that within the 
90\% confidence region for the MRST global fit, the valence quark CSV 
could be up to four times as large as that 
predicted by Rodionov {\it et al.} and 
it could even have the opposite sign.  

The magnitude of the allowed CSV effects obtained by MRST is consistent 
with uncertainties in phenomenological PDFs.  The total momentum carried 
by valence quarks is determined to within about 2\%.   
Inclusion of a valence quark CSV term changes the momentum carried by  
valence quarks in the neutron from those in the proton.  The 
total momentum carried by valence quarks is given by the second 
moment of the distribution, \EG the momentum carried by up valence 
quarks in the neutron 
\beast
 U^n_{\V} &\equiv& \int_0^1 x\,u^n_{\V}(x) \, dx \ . 
\eeast     
The 90\% confidence limit in the valence quark CSV terms corresponds 
to a variation of roughly 2\% of the momentum 
carried by valence quarks 
in the neutron, or just about the known experimental uncertainty  
in this quantity. 

At the level allowed by MRST, isospin violating PDFs are 
sufficiently large that, by themselves, they could account for 
the entire anomaly in the Weinberg angle suggested by the NuTeV 
experiment \cite{NuTeV,NuTeV2,Lon03,Lon03b}.  At present, 
this is the only single effect that appears capable of removing 
100\% of the NuTeV anomaly \cite{Lon05}; consequently, it is quite 
important that one be able to test the magnitude of parton 
isospin violation.     

The MRST group also searched for the presence of charge 
symmetry violation in the sea quark sector.  Again, they chose a 
specific form for sea quark CSV, dependent on a single parameter,   
\bea 
\bar{u}^n(x) &=& \bar{d}^p(x)\left[ 1 + \delta \right] \nonumber \\  
\bar{d}^n(x) &=& \bar{u}^p(x)\left[ 1 - \delta \right] 
\label{eq:seaCSV}
\eea  
With the form chosen, the total momentum carried by antiquarks in 
the neutron and proton are approximately equal.  

Perhaps surprisingly, evidence for sea quark CSV in the global fit 
was substantially stronger than that for valence quark CSV.  As shown in 
the top curve in Fig.\ \ref{Fig:chi2}, the best fit is obtained for 
$\delta = 0.08$, corresponding to 
an 8\% violation of charge symmetry in the nucleon 
sea. The $\chi^2$ corresponding to this value is substantially 
better than with no charge symmetry violation, primarily because of 
the improvement in the fit to the NMC $\mu-D$ DIS data \cite{Ama91,Arn97} 
when $\bar{u}^n$ is increased.  The fit to the E605 Drell-Yan data 
\cite{E605} is also substantially improved by the sea quark 
CSV term. 

As explained in the following sections, we have used the 
MRST CSV parton distributions to calculate the 
differences one could expect in observables for pion-induced Drell-Yan 
processes, and semi-inclusive charged pion production in electron-deuteron 
deep inelastic scattering.

\section{Pion-Induced Drell-Yan Processes and Parton Charge 
Symmetry \label{Sec:dyCSV}}

\begin{figure}
\includegraphics[width=2.0in]{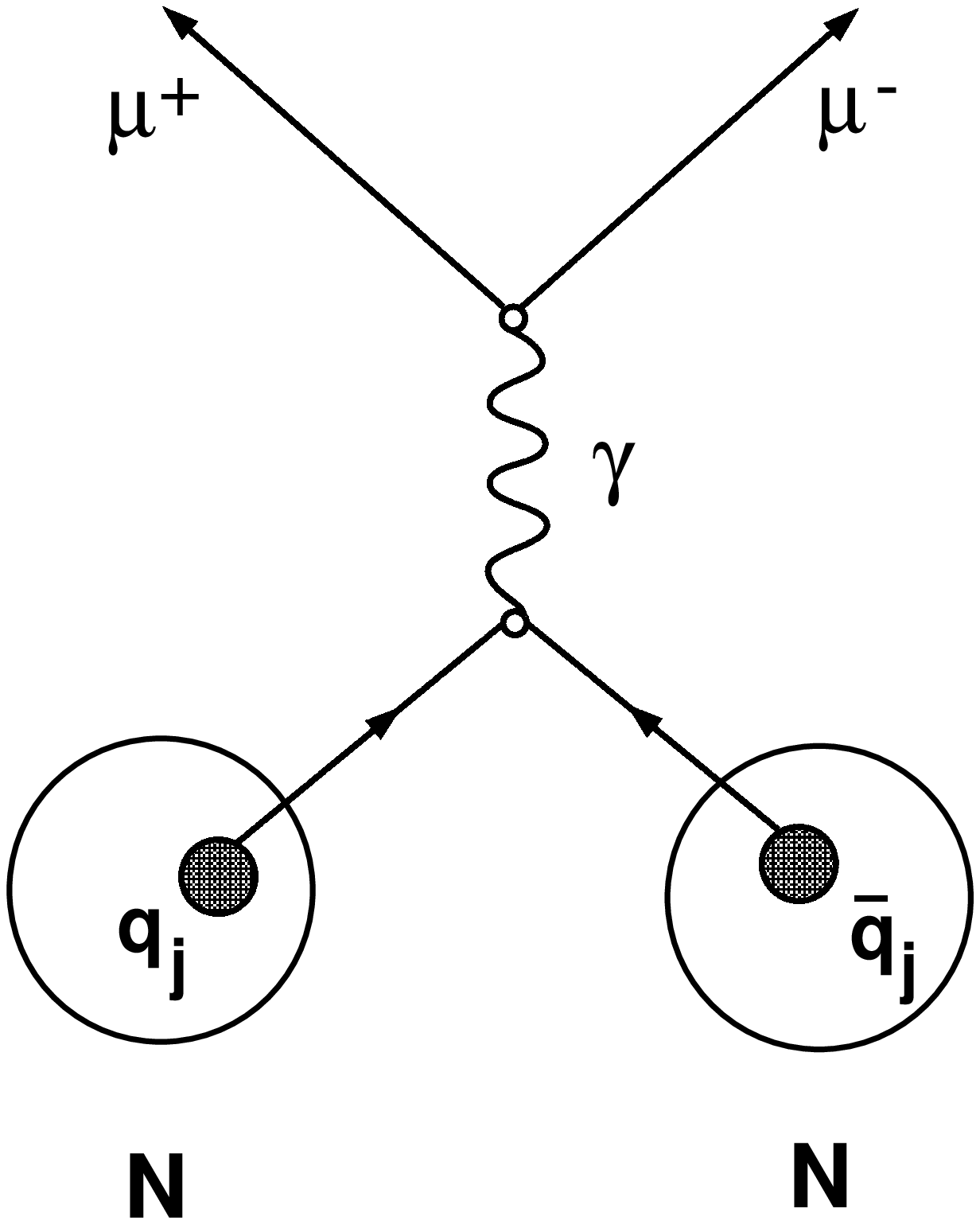}
\includegraphics[width=2.0in]{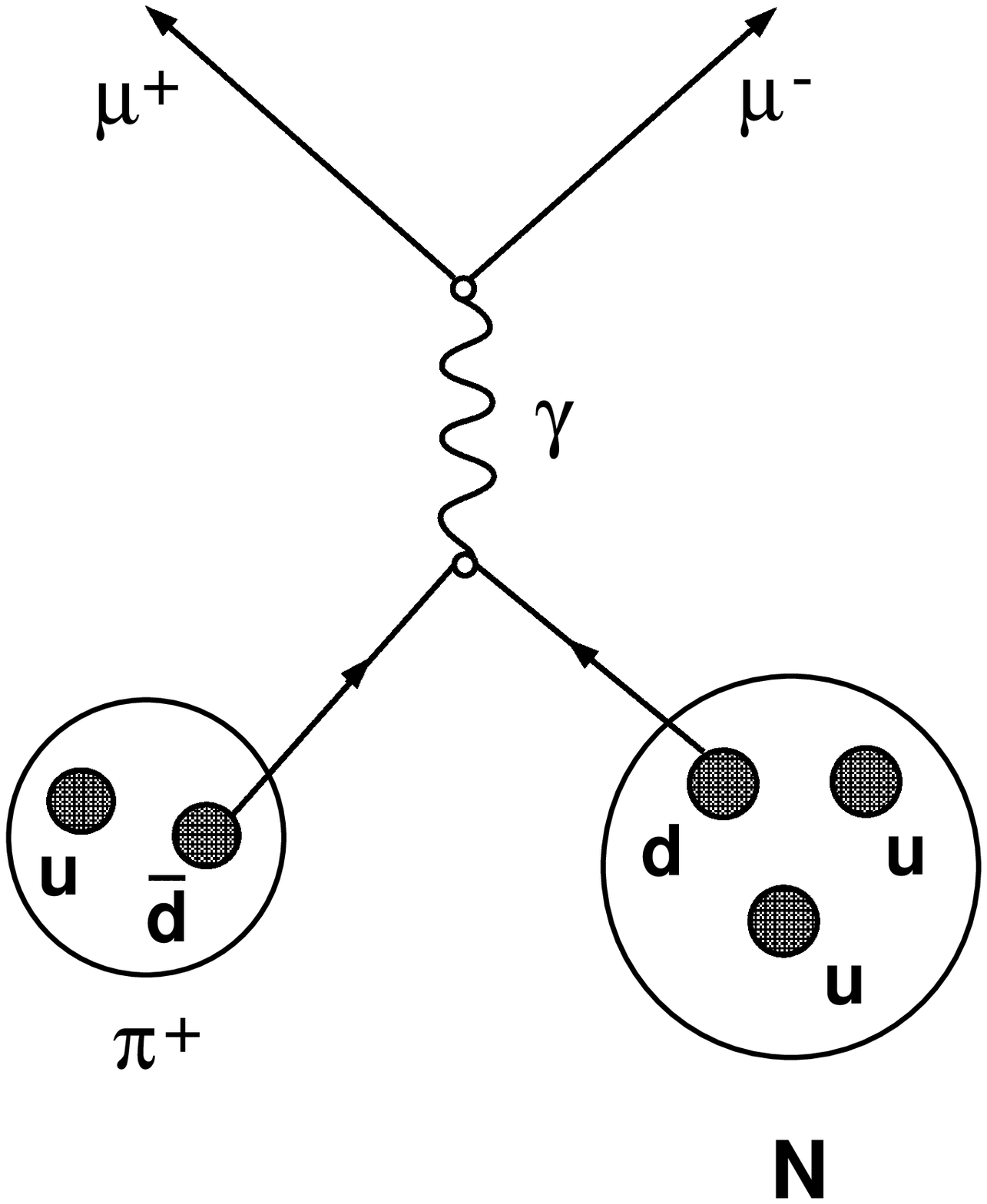}
\caption{Schematic picture of the Drell-Yan process; a quark and 
antiquark of the same flavor annihilate to form a virtual photon that 
decays into a high-mass muon pair. Left: In NN DY processes, a quark 
in one nucleon annihilates with an antiquark in the other nucleon. 
Right: in $\pi^+-p$ DY, a valence $\bar{d}$ in the $\pi^+$ can 
interact with a down quark in the proton. 
\label{Fig:dyan}}
\end{figure}

In Drell-Yan (DY) processes \cite{DY}, two hadrons collide at high energies, 
and a quark in one hadron annihilates an antiquark of the same flavor in the 
other hadron, producing a virtual photon which subsequently radiates a pair 
of muons with opposite sign.  This is shown schematically in Fig.\ 
\ref{Fig:dyan}, for $NN$ and $\pi N$ DY processes.  

Since the PDFs for nucleons are rather accurately known, DY processes 
for pions on nucleons provide a sensitive way of extracting parton 
distributions in the pion. For example, the NA10 experiment at CERN 
\cite{Bet85} and experiment E615 at FermiLab \cite{Con89} both 
studied Drell-Yan reactions induced by $\pi^-$.   
Such reactions could be studied using the Main 
Injector at FermiLab to produce high energy protons that are 
scattered from a nuclear target to produce charged pions, and subsequently 
observing Drell-Yan reactions for these pions on an isoscalar target, 
which we will assume is the deuteron.  If this program 
were carried out, then from the DY reaction the pion PDFs could be 
very accurately determined at this value of $Q^2$.  We will show that 
by comparing DY cross sections for $\pi^+$ and 
$\pi^-$ on the deuteron, one can extract the CSV violating 
part of the nucleon valence PDFs -- alternatively, through such 
reactions one could produce considerably  
stronger upper limits on the parton CSV terms.  We use the nucleon 
CSV distributions recently extracted by MRST to indicate the range of CSV 
that could be probed in such experiments.  

To test nucleon valence quark CSV, it is necessary to measure 
$\pi-D$ DY processes at kinematics corresponding to large $x$ for 
both pion and nucleon, \EG~$x, x_\pi \ge 0.3$.  In this region, 
where it is a reasonable first approximation to neglect 
sea quark effects, the Drell-Yan process will 
predominantly occur when a valence quark in the deuteron is  
annihilated by a 
valence antiquark in the pion.  Then the lowest order (LO) DY 
cross sections for $\pi^+-D$ and $\pi^--D$ have the approximate form,  
\bea
\sigma^{\DY}_{\pi^+\,D}  &\approx& {1\over 9}\,  
 \left[ d^p(x) + d^n(x)\right] \bar{d}^{\pi^+}(\xpi) \ ,  \nonumber \\ 
\sigma^{\DY}_{\pi^-\,D}  &\approx& {4\over 9}\,  
 \left[ u^p(x) + u^n(x)\right] \bar{u}^{\pi^-}(\xpi) \ .
\label{eq:piDY}
\eea
   
Now, from charge conjugation invariance and the assumption of charge 
symmetry for pion PDFs, we can write the relations, 
\bea 
 \pi_{\V}(x) &=& u_{\V}^{\pi^+}(x) = \bar{d}_{\V}^{\pi^+}(x) = 
  d_{\V}^{\pi^-}(x) = \bar{u}_{\V}^{\pi^-}(x) \ , \nonumber \\ 
 \pi_{\SC}(x) &=& q_{\SC}^{\pi^+}(x) = \bar{q}_{\SC}^{\pi^+}(x) = 
  q_{\SC}^{\pi^-}(x) = \bar{q}_{\SC}^{\pi^-}(x) \hspace{0.2in} 
 [q=u,d] \ , \nonumber \\ 
 \widetilde{\pi}_{\SC}(x) &=& s^{\pi^\pm}(x) = \bar{s}^{\pi^\pm}(x) \ . 
\label{eq:piPDFdef}
\eea
Inserting the results of Eq.~(\ref{eq:piPDFdef}) into Eq.~(\ref{eq:piDY}), 
we see that in the valence-dominated region for both pion and deuteron, the 
$\pi^--D$ DY cross section should be four times the $\pi^+-D$ 
term.  Thus, in the quantity 
\be
 4\sigma^{\DY}_{\pi^+\,D} - \sigma^{\DY}_{\pi^-\,D} \ , \nonumber 
\label{eq:DYcsv}
\ee  
the valence-valence contributions will cancel.  The remaining terms 
will contain sea-valence interference terms and charge symmetry 
violating terms for both nucleon and pion.  Introducing the nucleon CSV 
terms into the PDFs, the DY cross section for $\pi^+-D$ has the form, 
\bea
 9\sigma^{\DY}_{\pi^+\,D}(x_\pi,x) &=& \piv \left[ 
 \dvx + \uvx - \deluv + 5\left( \bux + \bdx \right) - \dub - 4\ddb \right] 
 \nonumber \\ &+& \pis \bigg[ 5\left(\dvx + \uvx \right) 
 + 10\left( \bux + \bdx \right) - \deluv - 4\deldv \nonumber \\ 
 &-& 2\dub - 8\ddb \bigg] + 2\pist \left[ s(x) + \bar{s}(x)\right] \ , 
\label{eq:piDYdef}
\eea 
in Eq.~(\ref{eq:piDYdef}) we have also introduced the charge symmetry 
violating PDFs from Eq.~(\ref{eq:CSVdef}).   

As proposed by Londergan \EA~\cite{Lon94}, once the DY cross 
sections for $\pi^+$ and $\pi^-$ on deuterium are measured, 
one can extract values for the pion valence and sea 
quark parton distributions.  Using those distributions, one can 
then focus on the region of large $x$ for both pion and nucleon.  
In this region, one can construct ratios of the DY cross sections, 
for example 
\be
 R^{DY}_{\pi D}(\xpi,x) = { 4\sigma^{\DY}_{\pi^+\,D} - 
 \sigma^{\DY}_{\pi^-\,D} \over \sigma^{\DY}_{\pi^-\,D} - 
 \sigma^{\DY}_{\pi^+\,D} }
\label{eq:RDY}
\ee   
As we will see, in order to extract CSV terms in this ratio, the 
large quantities in the two terms in the numerator of Eq.~(\ref{eq:RDY}) 
will very nearly cancel.  This requires that one be able to obtain very 
accurate relative normalization of DY cross sections for charged pions.  
This can be achieved by normalizing the charged pion 
cross sections to the $J/\psi$ peak.  

Inserting the DY cross sections from Eq.~(\ref{eq:piDYdef}) (and the 
analogous $\pi^--D$ term), one obtains 
\bea 
 R^{DY}_{\pi D}(\xpi,x) &\approx& {5\pis \over \piv } + C(\xpi)\left[  
  R_{\CS}(x) + R_{\SV}(x) \right] + R_{\SC}(\xpi,x) \nonumber \\ 
 R_{\CS}(x) &=& { 4( \deldv - \deluv)  \over 3\qvx } \nonumber \\ 
 R_{\SV}(x) &=& { 5(\bux + \bdx )- \dub - 4\ddb \over \qvx } \nonumber \\ 
 R_{\SC}(\xpi,x) &=&  {2 \widetilde{\pi}_s(\xpi)\over \piv}\,
 {(s(x) + \bar{s}(x) ) \over \qvx} \nonumber \\ 
 C(\xpi) &\equiv& \left( 1 + {2\pis \over \piv} \right) \nonumber \\ 
 \qvx &\equiv& \uvx + \dvx 
\label{eq:RSVcoef}
\eea  

Eq.~(\ref{eq:RSVcoef}) is calculated to lowest order in small terms.  
With the exception of the CSV terms, 
all of the parton distribution functions should be known (the pion 
PDFs at that $Q^2$ could be extracted from the same Drell-Yan 
experiment).  Our proposal is that one construct the ratio 
$R^{DY}_{\pi D}$ and measure its $x$ dependence for fixed 
$\xpi \ge 0.3$.  In this case the first term in Eq.~(\ref{eq:RSVcoef}) is a 
constant, and that constant will decrease as $\xpi$ increases. 
From the known nucleon and pion valence PDFs, one can predict the value 
of this ratio assuming charge symmetry.  Deviations of the ratio from 
this prediction would be evidence for CSV in nucleon parton distributions.  
The terms $R_{\CS}(x)$ in Eq.~(\ref{eq:RSVcoef}) contains contributions from 
valence quark CSV.  Although the CSV term is small, with increasing 
$x$ it will become a progressively larger fraction of the ratio, since 
all other terms are proportional to sea quark distributions that  
fall rapidly with increasing $x$.   

We used phenomenological parton distributions for nucleons and 
pions, that could be evolved to the $Q^2$ region of interest for 
Drell-Yan processes.  To obtain their CSV parton distributions, 
MRST started with the MRST2001 set of PDFs and varied these 
using a global fit, to obtain the best distributions including 
charge symmetry violation.  The resulting PDFs differ slightly 
from the best-fit MRST2001 distributions \cite{ThornePC}.  
The value of $\Lambda_{\QCD}$ and the gluon distributions are 
essentially identical to those in the MRST2001 set \cite{MRST01}.  
The input parameters for the nucleon PDFs, allowing for valence 
quark CSV, are 
\bea 
 x u_{\V}(x) &=& 0.129 x^{0.223} (1-x)^{3.31} \left[ 1 + 4.89x^{0.5} 
 + 69.86x \right] \nonumber \\     
 x d_{\V}(x) &=& 0.0163 x^{0.241} (1-x)^{3.75} \left[ 1 + 123.6x^{0.5} 
 + 76.04x \right] \nonumber \\     
 x S(x) &=& 0.215 x^{-0.269} (1-x)^{7.37} \left[ 1 + 3.34x^{0.5} 
 + 11.80x \right] 
\label{eq:MRSTval}
\eea
MRST define the quark sea from the function $S(x)$ using the relations, 
\bea 
 \bar{u}(x), \ \bar{d}(x), \ \bar{s}(x) &=& 0.2S - 0.5\Delta, 
 \hspace{0.5cm} 0.2S + 0.5\Delta, \hspace{0.5cm} 0.1S \nonumber \\ 
 x\Delta(x) &=& x(\bar{d}(x) - \bar{u}(x)) \nonumber \\ 
  &=& 1.195x^{1.24} (1-x)^{9.10} \left[ 1+ 14.05x - 45.52 x^2 \right] 
\label{eq:MRSTsea}
\eea
The starting scale for these PDFs is $Q_0^2 = 1$ GeV$^2$, and 
MRST provide interpolation matrices that allow one to evolve these 
PDFs to higher $Q^2$.  For the valence quark CSV we used the 
MRST form of Eq.~(\ref{eq:CSVmrst}), and we varied the overall 
parameter $\kappa$ between the range $-0.8$ and $+0.65$ 
corresponding to the 90\% confidence limit obtained by MRST.  
We have used the same MRST CSV distributions 
given in Eqs.~(\ref{eq:CSVmrst}) 
and (\ref{eq:seaCSV}) at all $Q^2$, in agreement with MRST, who 
did not include any $Q^2$ dependence in the CSV terms in their global 
fits.   

In the MRST global fits, the valence and sea quark CSV were varied 
separately, and slightly different minima were obtained for the 
quark PDFs in these two cases.  Thus there is some inconsistency in 
our using sea quark CSV terms with the quark PDFs appropriate for 
the valence CSV global fit.  However, the contribution of sea quark 
CSV to the ratio $R^{DY}_{\pi D}$ of Eq.~(\ref{eq:RDY}) is extremely 
small since we restrict our attention to the region where both 
$x$ and $\xpi$ are large.  In this region the inconsistency in our 
procedure has only a very small effect.  

For the pion PDFs we took the parton distributions of Sutton 
\EA~\cite{Sut92}, extracted from pion Drell-Yan and prompt photon 
experiments \cite{Bet85,Con89,WA70}.  
The Sutton analysis obtained different pion PDFs, depending upon 
the amount of the pion momentum assumed to be carried by the sea.  
We used those PDFs that correspond to the sea carrying 10\% of 
the pion momentum.  These PDFs are defined for a starting scale 
$Q_0^2 = 4$ GeV$^2$ and can be evolved upwards using interpolation 
matrices provided by the Durham group \cite{DurWeb}. For an actual 
$\pi-D$ Drell-Yan experiment, the pion PDFs would be extracted 
directly from the Drell-Yan data at that $Q^2$.   
  
\begin{figure}
\includegraphics[width=3.3in]{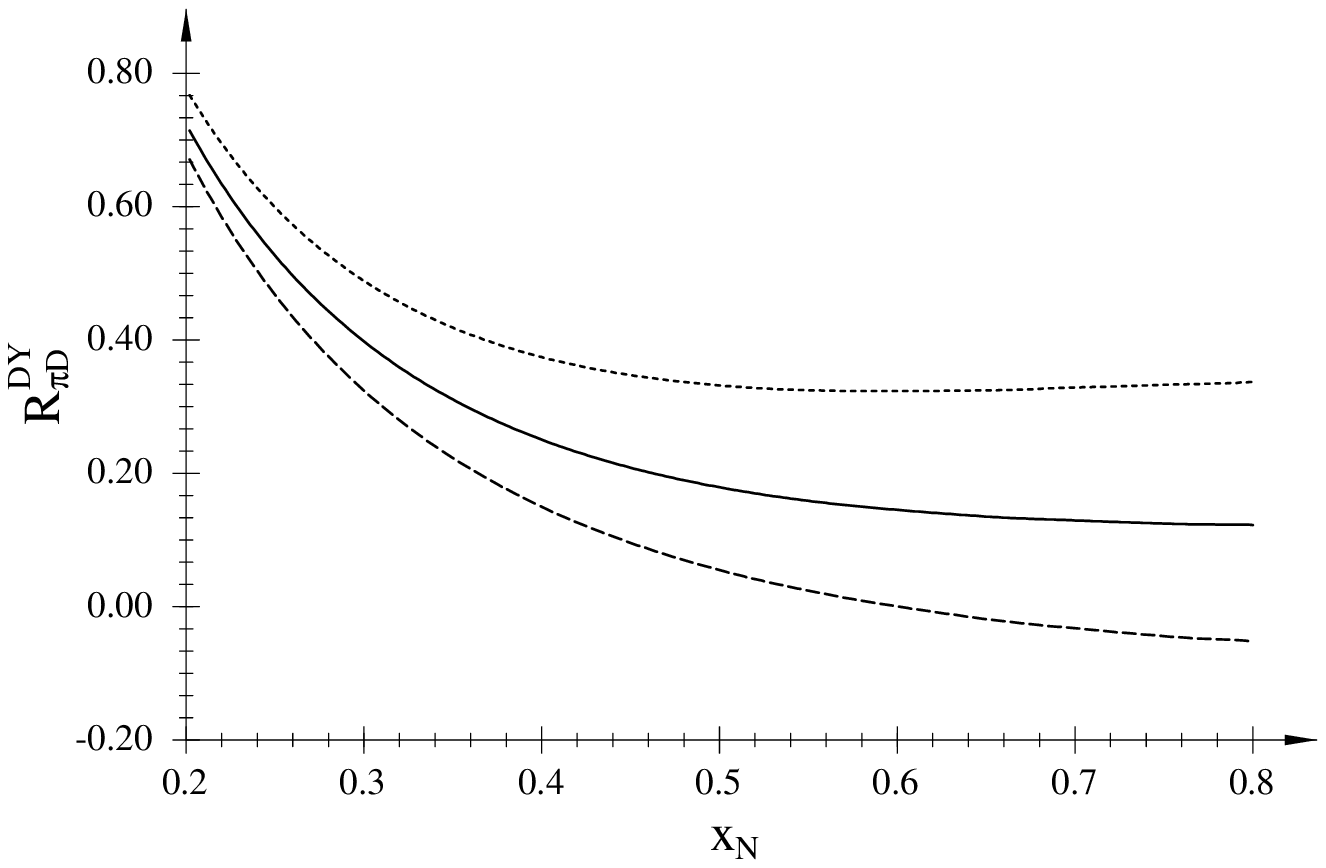}
\includegraphics[width=3.3in]{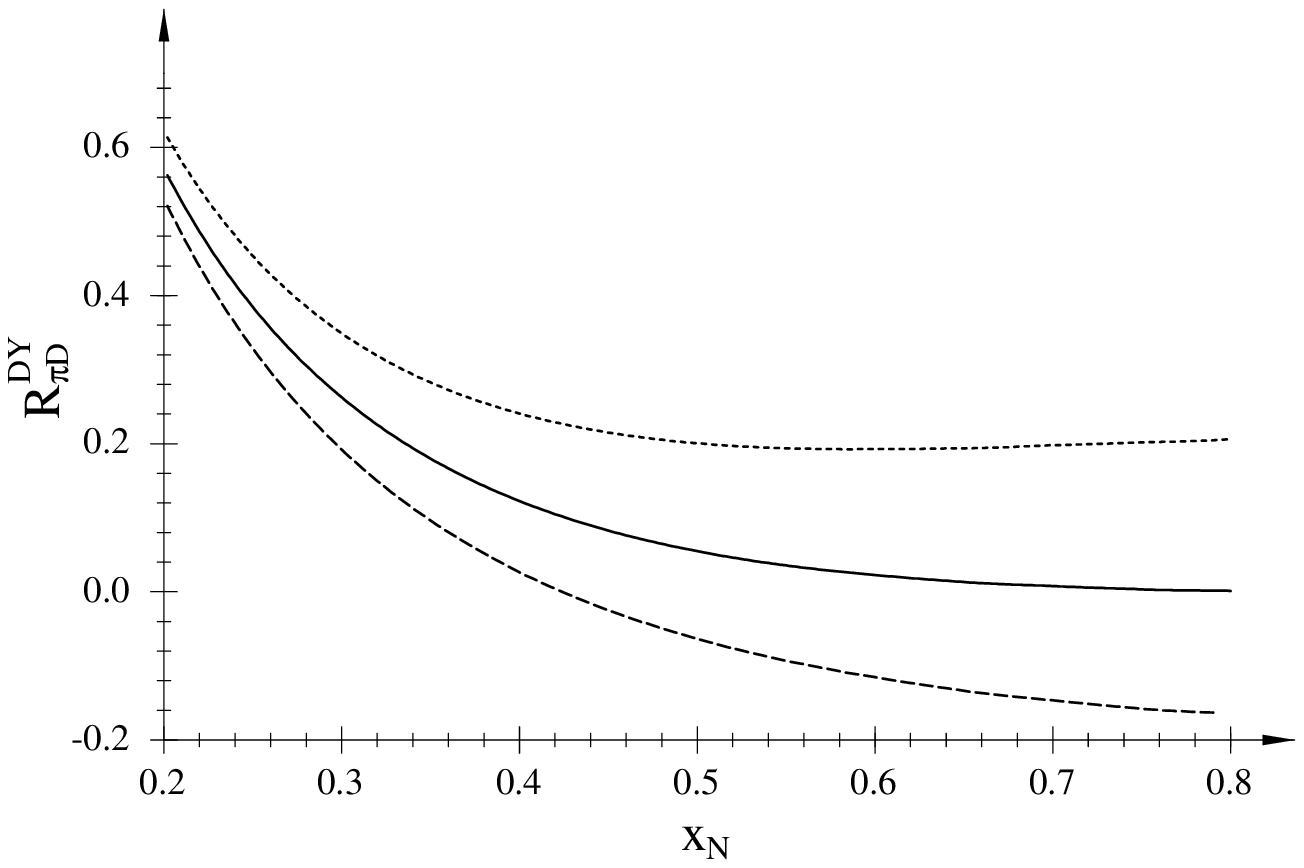}
\caption{The ratio $R^{DY}_{\pi D}$ defined in Eq.~(\protect\ref{eq:RSVcoef}) 
as a function of $x$, for $Q^2 = 25$ GeV$^2$. Top graph: $\xpi=0.4$; 
bottom graph: $\xpi=0.8$. Solid curve: no CSV terms, $\kappa = 0$; dashed 
curve: $\kappa = +0.65$; dotted curve: $\kappa = -0.8$.
\label{Fig:rsv25}}
\end{figure}

We have plotted the quantity $R^{DY}_{\pi D}$ defined in 
Eq.~(\ref{eq:RSVcoef}) as a function of $x$, for fixed $\xpi$,  
at $Q^2 = 25$ GeV$^2$, which is a representative value 
of $Q^2$ that could be obtained in fixed-target Drell-Yan experiments 
with charged pions at FermiLab.  In fixed-target experiments the 
detector configurations preferentially detect results corresponding 
to large $|x_F| = |x - x_{\pi}|$.  The top graph in Fig.~\ref{Fig:rsv25} 
plots $R^{DY}_{\pi D}$ vs.~$x$ for $\xpi = 0.4$, while the bottom 
graph shows the same quantity for $\xpi = 0.8$.     

The solid curve in Fig.~\ref{Fig:rsv25} corresponds to no CSV 
contribution; the dashed and dotted curves include the limits of the 
phenomenological CSV contributions at the 90\% confidence level as 
extracted by the MRST group.  The nucleon PDFs are extracted from 
global fits, and the pion PDFs would have previously been measured 
in this DY experiment.  The difference between the solid curve 
and the two dashed curves gives the magnitude of CSV effects 
(almost entirely nucleon valence CSV) allowed by the MRST phenomenological 
fit. Note that at reasonable kinematic values, the differences between 
the CSV contributions are substantial.  At $x = 0.4$ the two extremes of 
the CSV contributions are 30-50\% smaller or larger than the value 
corresponding to no CSV contribution, while for $x \sim 0.7$ the difference 
between the terms is more like a factor of 2. 

The magnitude of the solid curve gives the precision necessary 
in the DY measurements in order to extract meaningful information.  
For $\xpi = 0.4$, the two DY cross sections are predicted to cancel to 
within about 10\%.  Therefore the charged pion DY cross sections need to 
be measured to a few percent.  For $\xpi = 0.8$ and $x \sim 0.7$, the two DY 
cross sections cancel almost completely with the pion PDFs used in these 
calculations.  However, the CSV contributions at the 90\% level would allow 
values between roughly $-0.2$ and $+0.2$, so measurements of DY cross sections 
at the few percent level would allow one to discriminate between these 
limits.   

There is an additional contribution to the Drell-Yan ratio of 
Eq.~(\ref{eq:RSVcoef}) arising from the possibility of charge symmetry 
violation in the pion PDFs.  The largest contribution would arise 
from differences between the valence $\bar{d}$ in the $\pi^+$ and 
the $\bar{u}$ in the $\pi^-$.  The additional contribution 
to the DY ratio has the form 
\be
 \delta R^{DY}_{\pi D}(\xpi,x) \approx {4\left( \bar{d}_{\V}^{\pi^+}(x) - 
  \bar{u}_{\V}^{\pi^-}(x)\right)  \over 3\piv } 
\label{eq:piCSV}
\ee  
The contribution from pion CSV is a function only of $\xpi$.  For 
fixed $\xpi$ it will contribute a constant amount to the DY ratio.  
This was estimated by Londergan \EA~\cite{Lon94}, by taking  
quark mass effects into account in a Nambu-Jona Lasinio model that was used 
to calculate pion valence quark distributions.  The charge symmetry 
violating pion PDFs were calculated at a small value of $Q^2$ and then 
evaluated at higher $Q^2$ values using DGLAP evolution.  

The predicted effects from pion CSV were quite small.  For $\xpi \sim 
0.4$, the effects were almost zero, while for $\xpi \sim 0.8$ the 
contribution from Eq.~(\ref{eq:piCSV}) was less than 0.01 \cite{Lon94}.  
This term is added to all of the curves in Fig.~\ref{Fig:rsv25}, and 
increases slightly the uncertainty in nucleon CSV that can be 
extracted from these ratios.        

\section{Semi-Inclusive Electron-Deuteron Reactions and Parton Charge 
Symmetry Violation \label{Sec:SIDIS}}
  
Another possibility to measure charge symmetry violating terms in 
parton distributions arises in measurements of semi-inclusive 
charged pion production from lepton DIS on isoscalar nuclear targets.  
Considering semi-inclusive electroproduction of a hadron $h$ from 
a nucleon $N$, $e + N \rightarrow h + X$, the yield of hadron $h$ 
in such processes is given by 
\be 
 N^{Nh}(x,z) = \sum_i e_i^2 q_i^N(x)\,D_i^h(z) 
\label{eq:yield}
\ee
In Eq.~(\ref{eq:yield}), $N^{Nh}(x,z)$ is the yield of hadron $h$ 
from the semi-inclusive DIS (SIDIS) electroproduction on nucleon 
$N$, $q_i^N(x)$ is the parton distribution for flavor $i$ 
in the nucleon. 
Similarly $D_i^h(z)$ is the fragmentation function for a quark 
of flavor $i$ to fragment into hadron $h$, where $z$ is the 
fraction of the hadron energy carried by the quark.  Semi-inclusive 
production of a charged hadron from a proton thus has the form 
\bea
 N^{p\pm}(x,z) &=& {4\over 9}\left[ u^p(x)D_u^{\pm}(z) + 
  \bar{u}^p(x)D_{\bar{u}}^{\pm}(z) \right] \nonumber \\ 
 &+& {1\over 9}\left[ d^p(x)D_d^{\pm}(z) + 
  \bar{d}^p(x)D_{\bar{d}}^{\pm}(z) + s^p(x)D_s^{\pm}(z) + 
  \bar{s}^p(x)D_{\bar{s}}^{\pm}(z) \right] 
\label{eq:yieldpi}
\eea 
Charge conjugation invariance requires that 
\be
D_{\bar{u}}^{\pm}(z) = D_u^{\mp}(z), \hspace{0.5cm} 
  D_{\bar{d}}^{\pm}(z) = D_d^{\mp}(z) \ .
\label{eq:fragcsv}
\ee
The above equations hold for electroproduction of any charged hadrons. In the 
remaining discussion we look specifically at electroproduction of charged 
pions. If we assume charge symmetry for fragmentation functions we have 
the additional relations  
\be
D_d^{\pi^-}(z) = D_u^{\pi^+}(z), \hspace{0.5cm} 
  D_d^{\pi^+}(z) = D_u^{\pi^-}(z) \ .
\label{eq:cconj}
\ee
Thus, under the assumption of charge symmetry, the 
fragmentation of light quarks to charged pions can be written in 
terms of only two independent fragmentation functions, which are 
defined as ``favored'' or ``unfavored'' depending upon whether the quark 
that produces a hadron exists in the valence configurations of that hadron.  
Thus for up quarks $D_u^{\pi^+}$ is ``favored'' while $D_u^{\pi^-}$ is 
``unfavored''. 
     
From Eq.~(\ref{eq:yield}), the yields depend on ``favored'' and 
``unfavored'' fragmentation functions, as well as the fragmentation of 
strange quarks to pions.  Levelt, Mulders and Schreiber \cite{Lev91} derived 
an expression for measuring the ratio of fragmentation functions.  
They showed that 
\be
 {\langle N^{p\pi^+}(x,z) - N^{n\pi^+}(x,z) + N^{p\pi^-}(x,z) - 
  N^{n\pi^-}(x,z) \rangle \over \langle N^{p\pi^+}(x,z) - N^{n\pi^+}(x,z) 
  - N^{p\pi^-}(x,z) + N^{n\pi^-}(x,z) \rangle } = {9S_{\sst G}\over 5} 
  {\Delta(z) + 1 \over \Delta(z) - 1}
\label{eq:LMS}
\ee
In Eq.~(\ref{eq:LMS}), the brackets denote integration of both numerator 
and denominator over $x$, the quantity $\Delta(z)$ is the 
unfavored/favored ratio 
\be 
\Delta (z) \equiv {D_u^{\pi^-}(z) \over D_u^{\pi^+}(z)} \ ,  
\label{eq:Delta}
\ee
and $S_{\sst G}$ is the Gottfried sum rule \cite{Got67}.  Thus the 
semi-inclusive electroproduction of charged pions in $e-D$ reactions 
can be used to extract the unfavored/favored ratio of fragmentation 
functions of quarks to pions.  

Londergan \EA~\cite{Lon96} showed that these semi-inclusive 
reactions could also be used to investigate the presence of 
charge symmetry violation in nucleon parton distributions. This 
follows from the fact that, at large $x$ and assuming parton 
charge symmetry, the favored production 
of charged pions from valence quarks obeys the relation 
\be 
N^{D\pi^+}(x,z) \approx 4 N^{D\pi^-}(x,z)
\label{eq:favcs}
\ee
Consequently, they proposed measuring the ratio 
\be 
R(x,z) \equiv {4 N^{D\pi^-}(x,z)- N^{D\pi^+}(x,z)\over N^{D\pi^+}(x,z)
   - N^{D\pi^-}(x,z) }
\label{eq:Ratio}
\ee
It is convenient to define the quantity
\be 
R^D(x,z) \equiv {1- \Delta(z)\over 1 + \Delta(z)} R(x,z) \ . 
\label{eq:DRatio}
\ee
Note that the overall $z$-dependent factor in Eq.~(\ref{eq:DRatio}) is just 
the factor that is extracted using the Levelt \EA~ratio defined in 
Eq.~(\ref{eq:LMS}).  To lowest order in small quantities it can be shown that
\bea 
R^D(x,z) &=& R^D_f(z) + R_{\CS}(x) + R_{\SV}(x)+ R^D_{\SC}(x,z) \nonumber \\ 
R^D_f(z) &=& {5\Delta(z)\over 1 + \Delta(z)}  \nonumber \\ 
  R_{\SC}^D(x,z) &\equiv& {D_s^{\pi^+}(z) + D_s^{\pi^-}(z)\over D_u^{\pi^+}(z)
 (1 + \Delta(z))} {\left( s(x) +\bar{s}(x)\right) \over \qvx }\ .
\label{eq:RDdef}
\eea 
The quantities $R_{\CS}(x)$, $R_{\SV}(x)$ and $q_{\V}(x)$ are those 
defined in Eq.~(\ref{eq:RSVcoef}).  Thus the 
quantity $R^D(x,z)$ can be divided into three parts: a term that  
depends only on $z$; a term that contains the parton CSV 
contribution, and that depends only on $x$; and terms that contain 
contributions from sea quarks.  The term $R^D_f(z)$ depends on the 
unfavored/favored fragmentation function ratio, which can be accurately 
measured in these semi-inclusive reactions using the method proposed by 
Levelt \EA~(there is a contribution from charge symmetry violation in 
the fragmentation functions; this term was discussed in Ref.~\cite{Lon96}.  
It is expected to be small, and it depends only on $z$).  In order to 
isolate the CSV contribution, one should examine the ratio $R^D(x,z)$  
of Eqs.~(\ref{eq:Ratio}) and (\ref{eq:DRatio}) as a function of $x$ at 
fixed $z$.  Since the CSV 
term is expected to peak at $x \sim 0.35$ (see Fig.~\ref{Fig:MRSTfx}), and 
the sea quark contributions decrease very rapidly at large $x$, it is 
possible that at fixed $z$ and sufficiently large $x$, the CSV terms 
will be substantial, and might even dominate.  

For our calculations we used the fragmentation functions extracted by Kretzer, 
Leader and Christova \cite{Kre01}.  They used the information on $e^+e^-$ 
production on charged pions at the $Z^0$ peak \cite{Bin95}, together with 
the HERMES data on SIDIS charged pion production \cite{Air01}.  It 
turns out that the $e^+e^-$ data at the $Z^0$ peak essentially fixes 
the combination 
\be 
 D_{\Sigma}^{\pi^+} = 2\left[ D_u^{\pi^+} + D_d^{\pi^+} +D_s^{\pi^+} 
  \right] 
\label{eq:DSig}
\ee
This data, combined with the HERMES SIDIS data, allowed Kretzer {\it et al.} 
to extract the individual fragmentation functions, despite the 
fact that the $Z^0$ data need to be evolved over a great distance 
to match up with the HERMES data.  They obtained the relations 
\bea
D_u^{\pi^+}(z) &=& 0.689 z^{-1.039}(1-z)^{1.241} \nonumber \\   
D_d^{\pi^+}(z) &=& 0.217 z^{-1.805}(1-z)^{2.037} \nonumber \\   
D_s^{\pi^+}(z) &=& 0.164 z^{-1.927}(1-z)^{2.886} 
\label{eq:frags}
\eea
The fragmentation functions in Eq.~(\ref{eq:frags}) are appropriate 
for an average $\langle Q^2\rangle = 2.5$ GeV$^2$.  In this process, 
the fragmentation functions $D_u^{\pi^+}$ and $D_d^{\pi^+}$ are 
quite accurately determined, while the strange fragmentation function 
is determined to within about a factor of 2.    

\begin{figure}
\includegraphics[width=3.0in,angle=-90]{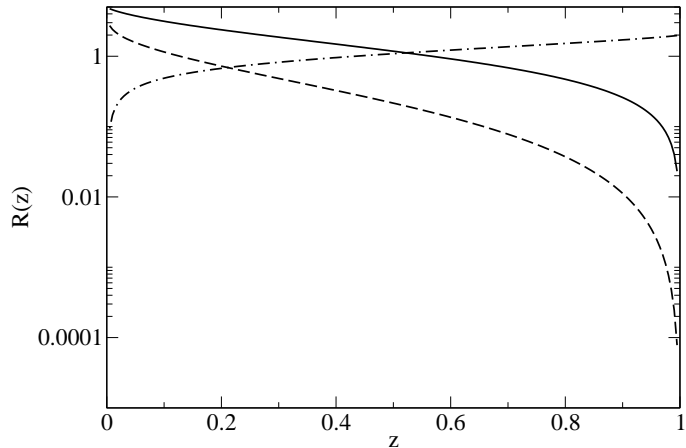}
\caption{Ratios of fragmentation functions vs.~$z$. Solid line: 
$R^D_f(z)$; dashed line: $z$-dependent factor in $R^D_{\SC}(x,z)$, defined in 
Eq.~(\protect{\ref{eq:RDdef}}). Dot-dashed line: the factor $C^{\Delta}(z)$ 
defined in Eq.~(\protect{\ref{eq:DelRatsep}}). Curves are calculated for 
$Q^2 = 2.5$ GeV$^2$.
\label{Fig:fragrat}}
\end{figure}

\begin{figure}
\includegraphics[width=3.0in,angle=-90]{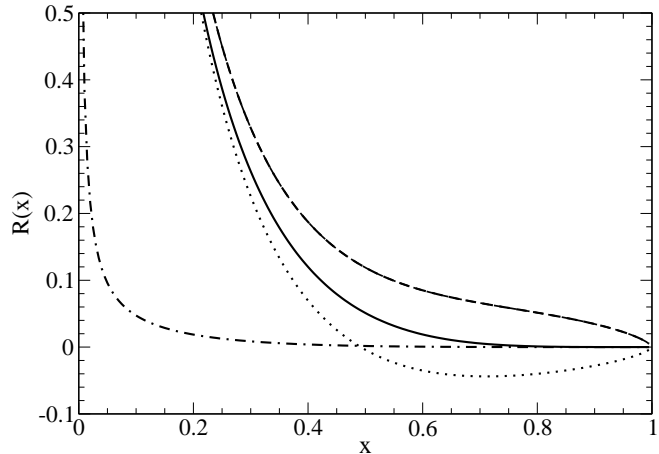}
\caption{Contributions of various terms to the ratio $R^D(x,z)$ defined 
in Eq.~(\protect{\ref{eq:DRatio}}) vs.\ $x$ at fixed $z=0.4$. Solid 
curve: sum of nonstrange and strange sea contributions, $R_{\SV}(x) + 
R^D_{\SC}(x,z=0.4)$. Dash-dot line: strange sea contribution, calculated 
for $z=0.4$. Long dash-dot (dotted) curves: inclusion of nucleon CSV 
terms from MRST global fit defined in Eq.~(\protect{\ref{eq:RSVcoef}}), 
for $\kappa = -0.8$ and $\kappa= +0.65$, respectively.  Curves are calculated 
for $Q^2 = 2.5$ GeV$^2$.
\label{Fig:sidisX}}
\end{figure}

In Fig.~\ref{Fig:fragrat}, we plot the ratios of fragmentation functions 
that enter into Eq.~(\ref{eq:RDdef}).  The solid curve is $R^D_f(z)$, 
while the dashed curve is the $z$-dependent factor in the term 
$R^D_{\SC}(x,z)$.  In Fig.~\ref{Fig:sidisX}, we plot the contributions to the 
quantity $R^D(x,z)$ vs.~$x$ for fixed $z=0.4$.  At this value, $R^D_f(z) \sim 
1.4$.  The solid curve corresponds to the case with no parton CSV 
contribution (this includes the sea and strange quark contributions).  
The dashed and dotted curves include valence parton CSV 
contributions corresponding to the MRST phenomenological PDFs with 
$\kappa = -0.8$ 
and $+0.65$, respectively.  These values demarcate the 90\% confidence 
limits using the MRST valence quark CSV function of Eq.~(\ref{eq:CSVmrst}). 
The dot-dashed curve is the 
strange contribution $R^D_{\SC}(x,z)$
calculated at $z=0.4$.  Except for extremely small values of $x$, the 
sea quark contribution is negligible. Hence the factor 2 uncertainty 
in the strange fragmentation function does not affect this ratio.   

For $x \ge 0.4$, the contributions from charge symmetry violating PDFs 
are substantial, and they rapidly become the dominant contribution at 
larger $x$.  Thus, at the levels determined by the MRST global fit, 
it would appear that precise measurements of charged pion production 
in semi-inclusive DIS electroproduction reactions on deuterium 
have the possibility of observing these isospin-violating effects. 
At the very least they would be able to lower the current 
allowed limits on partonic CSV effects.      

Note that the expected effects shown in Fig.~\ref{Fig:sidisX} are small.  
They would require extremely precise measurements, in order to be 
able to distinguish between various predictions for valence quark 
CSV terms.  In addition, Eq.~(\ref{eq:RDdef}) shows that the $x$-dependent 
terms plotted in Fig.~\ref{Fig:sidisX} sit atop an $x$-independent term.  
For $z = 0.4$ this term has a value of approximately 1.4; hence this 
term is much larger than the CSV term one wants to extract. This  
difficulty could be overcome, provided that one can measure very accurate 
values for the fragmentation functions. In this case one can construct 
the ratio 
\be 
R^{\Delta}(x,z) \equiv {8\left( {{\textstyle N^{D\pi^-}(x,z)}\over 
 {\textstyle 1 + 4\Delta(z)}} - 
 {{\textstyle N^{D\pi^+}(x,z)}\over {\textstyle 4 + \Delta(z)}}\right) 
  \over N^{D\pi^+}(x,z) - N^{D\pi^-}(x,z) }
\label{eq:DelRatio}
\ee
It is straightforward to show that 
\bea 
R^{\Delta}(x,z) &=& C^{\Delta}(z) \left[ R_{\CS}(x) + R_{\SV}(x)+ 
 R^D_{\SC}(x,z) \right] \nonumber \\ 
 C^{\Delta}(z) &=& {8(1 + \Delta(z)) \over (1 + 4\Delta(z))(4 + \Delta(z)) } 
\label{eq:DelRatsep}
\eea 
Eq.~(\ref{eq:DelRatio}) has the advantage that it eliminates the large 
$z$-dependent term in Eq.~(\ref{eq:RDdef}).  However, uncertainties in 
the favored/unfavored fragmentation ratio now play an important role in 
determining the uncertainty associated with extracting the charge symmetry 
violating PDFs.  The term $C^{\Delta}(z)$ in Eq.~(\ref{eq:DelRatsep}) 
is plotted as the dot-dashed curve in Fig.~\ref{Fig:fragrat}, at 
$Q^2 = 2.5$ GeV$^2$. It is normalized so that $C^{\Delta} \sim 1$ for 
moderate values of $z$.  

Using either of the ratios defined in Eqs.~(\ref{eq:RDdef}) or 
(\ref{eq:DelRatio}), in order to test for CSV terms in electron induced 
SIDIS reactions it is essential that one be able 
to vary $x$ while keeping $z$ constant.  Furthermore, the ability to write 
the ratio as the sum of terms in $x$ and $z$ requires that the 
cross section factorize as in Eq.~(\ref{eq:yield}).  Thus the data must be 
taken at sufficiently high energies that factorization is valid to within a 
few percent. SIDIS charged-pion experiments have been 
carried out at HERMES.  Another possibility is measurements of $e + D 
\rightarrow \pi^{\pm} + X$ at Jefferson Laboratory.  However, 6 GeV is most 
probably too low an energy for factorization to be valid, and even 12 GeV 
may not suffice.  It will be necessary to perform checks of the 
validity of the factorization hypothesis. If an electron-ion collider were 
built, these experiments could be carried out at sufficiently high energies.   

The CSV effects shown in Fig.~\ref{Fig:sidisX} are slightly underestimated. 
The relative magnitude of CSV and sea quark terms is the same for the 
Drell-Yan ratio and electroproduction, as can be seen by comparison of 
Eqs.~(\ref{eq:RSVcoef}) and (\ref{eq:RDdef}). The DY cross sections 
are evaluated at much larger $Q^2$ values; at these values the sea quark 
distribution is shifted to significantly smaller $x$.  However, we have not 
evolved the CSV distributions, in agreement with the procedure used by MRST 
in their global fit. If we evolved the CSV distributions in $Q^2$, this 
would increase the CSV contributions relative to the sea in 
Fig.~\ref{Fig:sidisX}.  Finally, our results have been derived in lowest order 
QCD; it is important to check how these results change in NLO.

\section{Conclusions \label{Sec:concl}}

The MRST phenomenological global fit to parton distribution functions 
provides limits on the magnitude of isospin violating PDFs.  The  
CSV effects allowed by the MRST fit are substantially larger than 
theoretical predictions of charge symmetry violation in parton 
distributions \cite{Sat92,Rod94}.  In this paper, we have analyzed 
two dedicated experiments that might detect parton CSV 
effects.  We used the range of values allowed by MRST to assess the 
magnitude of effects that might be expected in Drell-Yan reactions 
induced by charged pions on the deuteron, and in semi-inclusive 
production of charged pions in electron-deuteron deep inelastic 
scattering.  

In the Drell-Yan reaction one compares the cross sections for 
$\pi^+-D$ and $\pi^--D$ reactions at reasonably large values of $x$ for 
both pion and nucleon.  Using parton distributions for both nucleon 
and pion, we predict quite large values for this ratio, depending 
on the sign and magnitude of the CSV distributions.  

In the electron SIDIS measurements, once again one compares rates for 
production of $\pi^+$ and $\pi^-$.  In order to extract CSV terms in 
SIDIS reactions, it is necessary to make very precise measurements 
of both the cross sections and fragmentation functions. It is essential that 
the factorization 
hypothesis for the cross section be valid to a few percent. 
Nevertheless, given the importance of these quantities and their 
relevance in experiments like the NuTeV neutrino cross sections,  
it is of great interest now to investigate this 
issue experimentally.  

\section*{Acknowledgments\label{Sec:Acknwl}}

This work was supported in part [JTL] by National
Science Foundation research contract PHY0302248 and [AWT] by  
DOE contract DE-AC05-84ER40150, under which SURA operates Jefferson 
Laboratory. The authors wish to thank K. Hafidi, P. Reimer and R.S. 
Thorne for useful discussions regarding the issues presented here.

\end{document}